\documentclass[10pt,aps,twocolumn,showpacs,pra]{revtex4}
\usepackage{amssymb,amsfonts,amsmath,wrapfig,mathrsfs,mathbbol}
\usepackage{graphicx}

\begin{document}

\title{Dipole spectrum structure of non-resonant non-pertubative driven two-level atoms}

\author{A. Pic\'{o}n$^{1,3}$,  L. Roso$^{2}$, J. Mompart$^{3}$, O. Varela$^{4}$, V. Ahufinger$^{3,5}$, R. Corbal\'an$^{3}$, L. Plaja$^{4}$}

\affiliation{$^{1}$JILA, University of Colorado, Boulder 80309-0440, USA (present address)}

\affiliation{$^{2}$Centro de L\'aseres Pulsados (CLPU), E-37008 Salamanca, Spain}

\affiliation{$^{3}$Grup d'\`{O}ptica, Universitat Aut\`{o}noma de Barcelona, E-08193 Bellaterra (Barcelona), Spain}

\affiliation{$^{4}$Departamento de F\'\i sica Aplicada, Universidad de Salamanca, E-37008 Salamanca, Spain}

\affiliation{$^{5}$ICREA. Instituci\'o Catalana de Recerca i Estudis Avan\c{c}ats, Llu\'is Companys, 23. E-08010 Barcelona, Spain}

\date{\today}

\begin{abstract}
We analize the dipole spectrum of a two-level atom excited by a non-resonant intense monochromatic field, under the electric dipole approximation and beyond the rotating wave approximation. We show that the apparently complex spectral structure can be completely described by two families: harmonic frequencies of the driving field and field-induced nonlinear fluorescence. Our formulation of the problem provides quantitative laws for the most relevant spectral features: harmonic ratios and phases, non-perturbative Stark shift, and frequency limits of the harmonic plateau. In particular, we demonstrate the locking of the harmonic phases at the wings of the plateau opening the possibility of ultra-short pulse generation through harmonic filtering.
\end{abstract}

\maketitle

\section{Introduction}

Two-level atoms are essential blocks for the understanding of basic processes in quantum physics. In particular they are the keystone of quantum optics as the harmonic oscillator is the basic model for classical optics.  Their simplicity brings the possibility for analytical developments which convey physical interpretations not attainable with more exact approaches. Moreover, two-level models have good quantitative accuracy in the description of laser-matter interactions near resonance at moderate laser intensities, where the rotating wave approximation is applied  \cite{scully}. For the case of strong fields, where a large number of atomic or molecular transitions are simultaneously involved and ionization is present, the two-level model does not provide a complete description, although it can still be used to study the role of bound state transitions in high-order harmonic generation  \cite{conej98A}. Nevertheless, under particular conditions some of these complex systems are dominated by two-level transitions even for the strong driving case (for instance, the molecular Hydrogen ion at large internuclear distances \cite{zuo_93A}). Two-level transitions have also been associated to harmonic generation in molecules involving a rescattering center different from the electron's parent ion \cite{kopol98A} and connected with charge-resonant states in odd-charge molecular ions \cite{ivano93A}. Nowadays  the recent trends in OP-CPA (Optical Parametric Chirped-Pulse Amplification) techniques to produce high-power mid-infrared laser radiation \cite{ghotb06A} renew the interest of few-level systems in the off-resonant strong-coupling regime, since in this spectral region few atomic transitions can be reasonably isolated from the rest \cite{matis95A, biegert09}.  In addition, the forthcoming development of extreme intensity laser sources  may permit to directly address the excitation of dipolar two-level transitions in nuclei \cite{keite06A}.

Two-level atoms interacting with intense laser light show a significant non-linear behavior dominating the complex temporal dynamics. This dynamics is translated to the atom dipole evolution and, therefore, to the structure of the emitted field. The radiation spectra under such circumstances shows a plateau structure, in which harmonics with similar intensities extend to high frequencies \cite{sunda92A}. This is a universal behavior, present in other models of light-matter interaction (also in classical anharmonic oscillators) and the signature of the failure of the perturbative regime. The physics beneath this structure is well understood for the case of atoms and molecules in strong electromagnetic ionizing fields \cite{brabe00A}. In this case, the relevant process responsible for the plateau consists in the ionization of the electrons, their acceleration through the field-induced quiver motion, followed by the recollision with the parent ion with the release of the electron's kinetic energy into electromagnetic radiation.  On the other hand, for a two-level system, the  plateau emerges from transitions between adiabatic states as was pointed out in \cite{conej96A,figue02A} (see also the dicussion in the last section of this paper).

The study of coherently driven two-level systems beyond the perturbative limit has a fundamental reference in \cite{autle55A}, with special focus on the near resonant conditions. In Ref. \cite{plaja92A}, a quantitative description of the harmonic spectra of strongly driven two-level systems was reported by means of a Floquet-based theory. This approach resulted in a closed iterative analytical formulation, based on a continued-fraction method, that provided for exact ratios between the harmonic intensities. Later on, this approach was used to study the harmonic enhancement connected with field-induced multiphoton resonances \cite{plaja93A}. Other continued fraction approaches  have also been used to derive approximated analytical expressions for the plateau extension \cite{kapla94A}. The non-harmonic peaks also present in the radiative spectrum of the electric dipole have been studied for the case of large frequencies and/or very high amplitudes of the driving fields \cite{ivano93B,marti05A}. In these studies, the non-harmonic resonances are identified as satellite peaks around each harmonic, separated by the transition frequency between the quasienergy states. Two groups of peaks were identified (hyperRaman lines of Stokes and antiStokes type) according to their relative position with respect to the harmonic peaks. Also, these authors derived analytical formulas for the quasi-energies.  In the present paper we use the exact approach developed in  \cite{plaja92A} to address  the complete description of the dipole spectrum in the general case.  We will show that the apparent complexity of the spectra can be reduced to two contributions, namely, harmonic radiation and laser-induced fluorescence. This latter family accounts for the emergence of the mentioned satellite structure (Stokes and antiStokes lines) in the case of strong driving and/or in the limit of high driving frequencies. In particular, our approach gives a simple explanation of the asymmetry between the intensities of the Stokes and antiStokes satellite peaks. On the other hand, we obtain a polynomial equation whose solution gives the position of the Stark-shifted transition energy to, in principle, arbitrary accuracy. Restricting to the lowest degree of this equation, we derive an analytical formula for this later quantity. We test it against the exact numerical integration of the two-level Schr\"odinger equation, and show the convergence to the result  of  \cite{marti05A} (in the limit of validity considered in this reference). We also derive approximated formulas for the extension of the harmonic plateau. Our formula for the cut-off energy agrees with  \cite{kapla94A} and converges to \cite{ivano93B} in the limit of Rabi frequencies much greater than the transition frequency. On the other hand, our formula for the plateau's on-set frequency converges to  \cite{kapla94A} in the low frequency limit and gives better comparison with the exact results for higher driving frequencies. Finally, from the phase of the higher-order harmonics, we show their locking and demonstrate numerically the possibility of producing attosecond pulses.

\section{Theory}

\begin{figure}[htbp]
\centering\includegraphics[width=10cm]{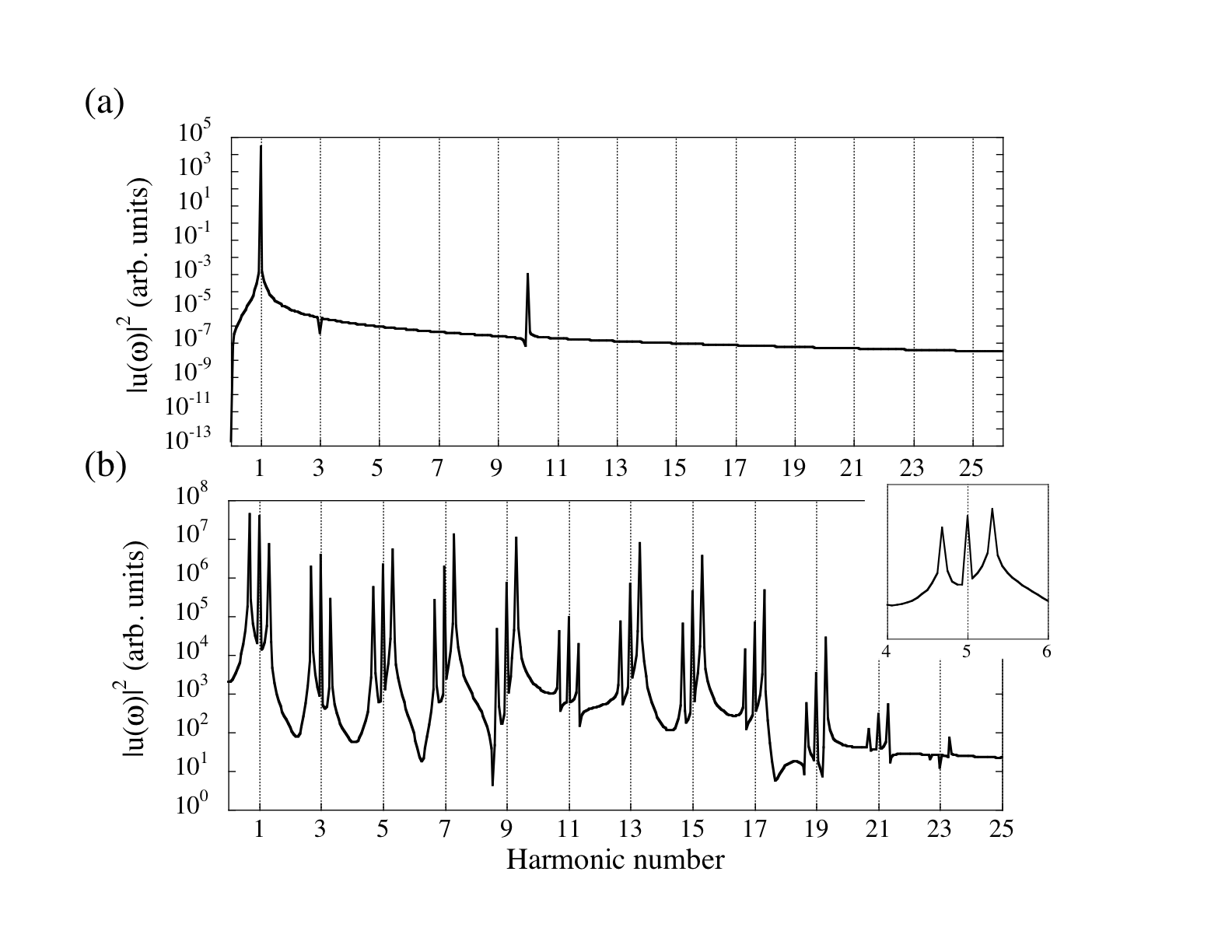}
\caption{\small Dipole spectrum, $|u(\omega)|^2$, of the driven two-level atom. (a) spectrum at low field amplitude ($\vert \chi \vert / \omega_{0} = 0.005$ and $\omega/\omega_0=0.1$), (b) spectrum at higher field amplitude ($\vert \chi \vert / \omega_{0} = 1.5$ and  $\omega/\omega_0=0.2$). Inset: detail of the dipole spectrum in (b) showing the structure of satellite peaks around the harmonics.}
\label{fig:two_spectra}
\end{figure}

We begin with the standard formulation for a two-level system driven by a monochromatic field, out of the rotating wave approximation. The dynamics of the probability amplitudes are given by:
\begin{eqnarray}
i  {d \over d t} a_1(t) & = & - {1 \over 2} \left[  \chi e^{-i \omega t} + \chi^* e^{i \omega t} \right] a_2(t) \label{eq:schrodinger1} \; , \\
i  {d \over d t} a_2(t) & = & \omega_0 a_2(t) - {1 \over 2} \left[  \chi e^{-i \omega t} + \chi^* e^{i \omega t} \right]  a_1(t) \; ,
\label{eq:schrodinger2}
\end{eqnarray}
where labels $1$ and $2$ stand for the lower and upper level respectively, $\omega_0$ is the atomic transition frequency, $\omega$ is the monochromatic field frequency, and $\chi=\mathcal{E}_0  \mu/\hbar$ is the Rabi frequency, where $\mathcal{E}_{0}$ is the electromagnetic amplitude and $\mu$ the dipole matrix element. The electromagnetic amplitude is defined by a linearly polarized monochromatic field of the form ${\bf E}(t)={\bf E}_0 \cos(\omega t + \phi)$, being ${\bf E}_0$ and $\phi$ the amplitude and the phase of the field respectively, with $\mathcal{E}_0=| {\bf E}_0 | e^{-i \phi}$, and the dipole matrix element as $\mu=\langle 1 | e z | 2 \rangle$, being a real quantity and assuming $z$ the field polarization direction. We define the Bloch variables $u$ and $v$ as the real and imaginary parts of $a_1(t)a_2^*(t)$, respectively, and the population inversion as $w(t)=|a_2(t)|^2-|a_1(t)|^2$. Then, from Eqs. (\ref{eq:schrodinger1}) and  (\ref{eq:schrodinger2}) it can be derived:
\begin{eqnarray}
{d \over d t} u(t) & = & - \omega_0 v(t) \; , \label{eq:blochu} \\
{d \over d t} v(t) & = &  \omega_0 u(t) + {1 \over 2}  \left[  \chi e^{-i \omega t} + \chi^* e^{i \omega t} \right] w(t) \; , \label{eq:blochv} \\
{d \over d t} w(t) & = &  -2  \left[  \chi e^{-i \omega t} + \chi^* e^{i \omega t} \right] v(t) \; .  \label{eq:blochw}
\end{eqnarray}
 The atomic dipole is defined as $d(t)\equiv2\mu u(t)$. Figure \ref{fig:two_spectra} shows the spectral content $\vert u(\omega)\vert^{2}$ computed numerically from Eqs. (\ref{eq:blochu})-(\ref{eq:blochw}). To be specific, in this paper we have chosen the driving field phase as $\phi= -\pi/2$. To reduce the effect of the abrupt envelope, we have considered a sinus squared field turn-on of two cycles, followed by $16$ cycles of constant amplitude. The dipole spectrum has been calculated through the Fourier transform of $u(t)$ in these later $16$ cycles.  For small field amplitudes, figure  \ref{fig:two_spectra}(a), the spectrum is dominated by two peaks, one located at the driving field frequency $\omega$ (Rayleigh scattering) and another at the transition frequency $\omega_0$ (field-induced fluorescence). As the field amplitude increases, figure \ref{fig:two_spectra}(b), the complexity of the dipole spectrum increases with the appearance of harmonic frequencies of the driving field surrounded by satellite peaks, see inset of figure \ref{fig:two_spectra}(b), whose position varies with the driving field intensity. These satellite structures are referred as hyperRaman lines in \cite{ivano93B,marti05A}. We should note that the relative intensity of the two satellite peaks around each harmonic varies with the driving field parameters,  the satellite structure being asymmetric in the general case. Also, note that the  harmonic structure shows the emergence of a plateau region with peaks of similar intensities. Mid-Infrared radiation can provide a suitable scenario in order to match the parameters considered in Fig. \ref{fig:two_spectra}(b). For example, the 2s-3p transition in the Hydrogen atom is close to 5 photon resonance using a 3 micron wavelength laser. In this case, the current technology is almost ready, see for example Ref. \cite{biegert09}.

Combining Eqs. (\ref{eq:blochu}) and (\ref{eq:blochv}) we obtain
\begin{equation}
{d^2 \over d t^2} u(t) + \omega_0^2 u(t) = - {\omega_0 \over 2}  \left[  \chi e^{-i \omega t} + \chi^* e^{i \omega t} \right]  w(t) \; ,
\label{eq:anharmonic}
\end{equation}
which describes the atomic dipole evolution. As discussed in \cite{plaja92A}, the dipole dynamics corresponds to the harmonic oscilator in the small coupling limit ($w(t) \simeq -1$) and, therefore, eq. (\ref{eq:anharmonic}) provides a connection between the fundamental atomic models of classical and quantum optics. On the other hand, combining Eqs. (\ref{eq:blochu}) and (\ref{eq:blochw}), we have
\begin{equation}
 {d \over d t} w(t) = {2 \over \omega_0}   \left[  \chi e^{-i \omega t} + \chi^* e^{i \omega t} \right]   {d \over d t} u(t) \; ,
 \label{eq:dwdu}
\end{equation}
an intermediate expression that we will use in the following. We now consider the spectral decompositions of $u$ and $w$,
\begin{eqnarray}
u(t) & = & \int_{-\infty}^\infty u(\alpha) e^{-i\alpha t} d \alpha \; , \\ 
w(t) & = & \int_{-\infty}^\infty w(\alpha) e^{-i\alpha t} d \alpha   \; .
\end{eqnarray}
Note that we are now considering Fourier components of arbitrary frequency, instead of the harmonic series in \cite{plaja92A}, restricted to integer multiples of the laser frequency.   The spectral forms associated with Eqs. (\ref{eq:anharmonic}) and (\ref{eq:dwdu}) are
\begin{eqnarray} \nonumber
u(\alpha) & = & - {\omega_0 \over 2} \left [ {\chi \over {\omega_0^2-\alpha^2} w(\alpha-\omega)} + {\chi^* \over {\omega_0^2-\alpha^2} w(\alpha+\omega)} \right ] \; , \\ \label{eq:u}\\
 w(\alpha) & = & {2 \over \omega_0} \left [ \chi {{\alpha - \omega} \over \alpha} u (\alpha-\omega) + \chi^* {{\alpha+\omega} \over \alpha} u(\alpha+\omega) \right] \; ,\nonumber \\ \label{eq:w}
\end{eqnarray}
respectively.


\section{The existence of two spectral families}
The combination of the later expressions (\ref{eq:u}) and (\ref{eq:w}) leads to a set of (infinite) coupled algebraic equations
\begin{equation}
\Phi_-(\alpha) u(\alpha-2\omega) + \Theta(\alpha) u(\alpha) + \Phi_+(\alpha) u(\alpha+2\omega)=0 \; ,
\label{eq:system}
\end{equation}
 with
 \begin{eqnarray}
 \Theta(\alpha) &=& 2 \alpha^2- {{(\alpha^2-\omega_0^2)(\alpha^2-\omega^2)}\over |\chi|^2}  \; ,\label{eq:Theta} \\
 \Phi_{\pm}(\alpha) &=& e^{\pm 2i \phi} (\alpha \pm 2 \omega) (\alpha \mp \omega) \; .
 \label{eq:Phi}
 \end{eqnarray}
Note from these equations that the dipole spectrum is partitioned in a set of independent families of the sort $\{ \alpha_0, \alpha_0 \pm 2 \omega, \alpha_0 \pm 4 \omega, \cdots \}$ (For instance, the set of harmonic peaks corresponds to a single family with $\alpha_0=\omega$).  This is a general result for two level systems interacting with a monochromatic field monochromatic fields, since up to this point we have not made any other approximation. 

Each spectral family is described by the following tri-diagonal matrix, generated by Eq. (\ref{eq:system}),
\begin{widetext}
\begin{equation} M(\alpha)=
\left( \begin{array}{cccccccc}
\ddots &  &  &  &  &  & &\\
  & \Theta(\alpha-4\omega) &   \Phi_{+}(\alpha-4\omega) &  &  & & &\\
  & \Phi_{-}(\alpha-2\omega)&  \Theta(\alpha-2\omega) &  \Phi_{+}(\alpha-2\omega) & & & &\\
  &  & \Phi_{-}(\alpha)&  \Theta(\alpha)&   \Phi_{+}(\alpha)&  & &\\
  &  &  & \Phi_{-}(\alpha+2\omega)  & \Theta(\alpha+2\omega)  & \Phi_{+}(\alpha+2\omega) & &\\
  &  &  &  & \Phi_{-} (\alpha+4\omega) &  \Theta(\alpha+4\omega) & &\\
  &  &  &  &  & & \ddots
\end{array} \right) \; . \label{eq:Matrix}
\end{equation}
\end{widetext}
Since Eq. (\ref{eq:system}) is an homogeneous system of equations, the necessary condition for the existence of a family of spectral peaks is $\det(M)=0$, yielding a polynomial equation (in principle of infinite degree), which gives exactly all the non-zero components of the dipole spectrum. The complete spectrum is, consequently, composed by all the families which are solutions of this equation. To find them we should address the general problem of finding all $\alpha$ so that $\det(M)=0$ is fulfilled. 

Since any physical spectrum cannot have Fourier components of arbitrarily large frequencies, all the spectral families must have some central region where the peaks have  relevant intensities. In the following, we shall consider $\alpha_0$ as a reference frequency  belonging to this central region. Away of this region, the intensities of the spectral peaks should decay gradually as their frequency approaches $\pm \infty$. This permits the truncation of $M$ at arbitrary large frequencies with almost perfect accuracy. Therefore, in the following, we shall consider  $M$  as truncated to dimension $ n \times n$ ($n$ odd) around the central frequency $\alpha_0$.  In this case, each family of spectral peaks is composed by $n$ components: $\{ \alpha_0, \alpha_0 \pm 2 \omega, \alpha_0 \pm 4 \omega, \cdots, \alpha_0 \pm (n-1)/2 \omega  \}$, each being a solution of $\det(M)=0$. Using definitions (\ref{eq:Theta}) and (\ref{eq:Phi}), the truncated $\det\left[M(\alpha) \right]=0$  leads to an algebraic equation of degree $4 n$ in $\alpha$. Therefore, the general spectrum will be composed at most by four families of peaks. Since the dipole is a real quantity, any solution for $\alpha$ implies a solution for $-\alpha$ (actually the associated Fourier component for $-\alpha$ is the complex conjugate of the one for $\alpha$). Thus there are only two independent non-vanishing families composing the spectrum. As we have pointed out before, the harmonic field is already one of such families, therefore there is only room for one more. We shall see later that this second family is associated with the natural frequency of the atom and, therefore, it is generated by the non-linear mixing of the two-level transition frequency with the laser frequency. Note this is also a general result for monochromatic fields.

\begin{figure}[htbp]
\centering\includegraphics[scale=0.33]{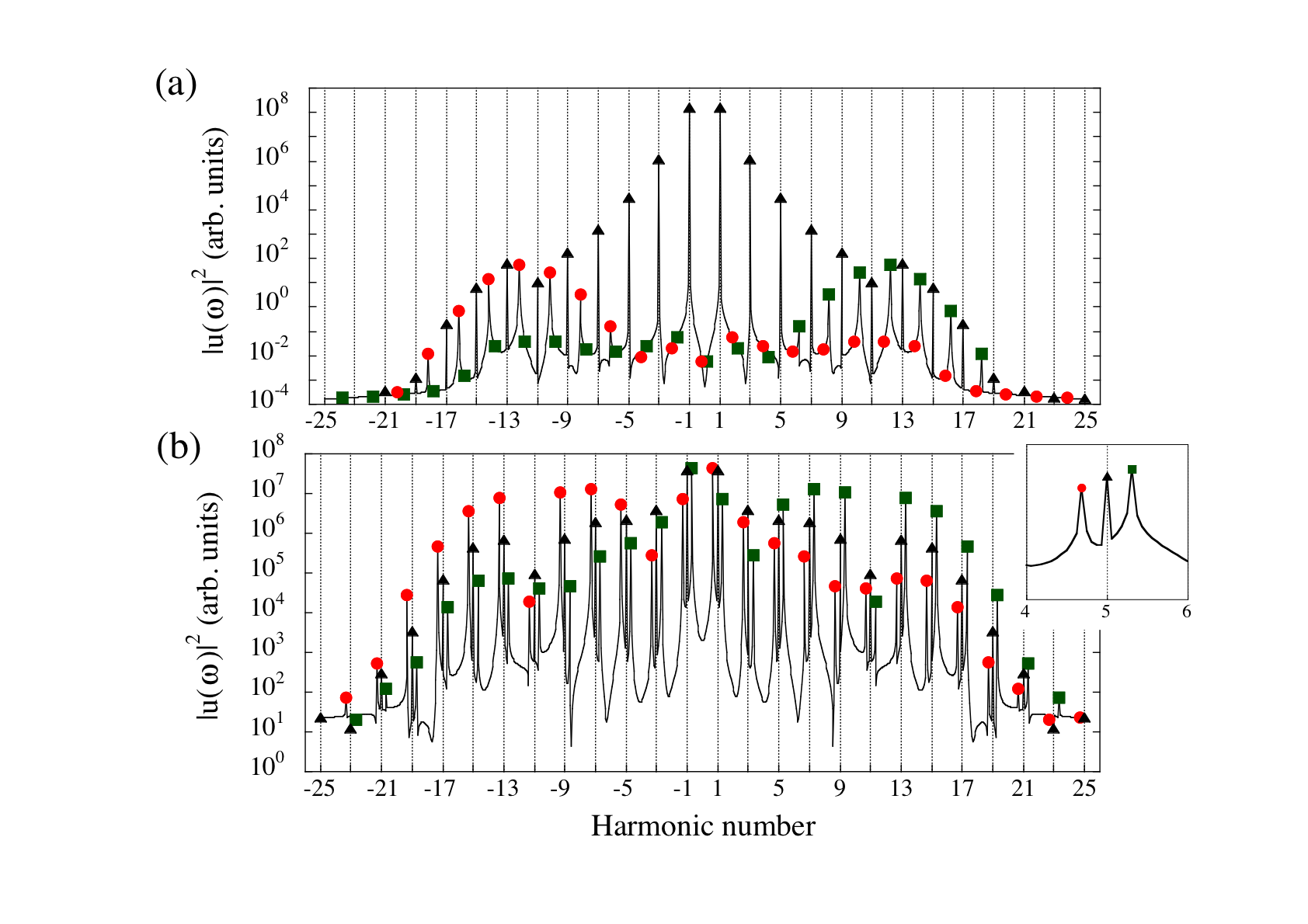}
\caption{\small Dipole spectrum, $|u(\omega)|^2$, plotted for negative and positive frequencies in two different cases (a)  $\omega/\omega_0=0.1$, $|\chi|/\omega_0=0.5$ , and (b) $\omega/\omega_0=0.2$, $|\chi|/\omega_0=1.5$. This later corresponds to the same case as in Fig. \ref{fig:two_spectra}(b), including the inset.  The two families composing the spectrum have been labeled with solid black triangles(harmonic family) and solid green boxes (fluorescence family). The complex conjugate of this later is labeled with solid red circles. }
\label{fig:full_spectra}
\end{figure}

 Figure \ref{fig:full_spectra} shows the exact solution of the dipole spectrum computed from Eqs. (\ref{eq:schrodinger1}) and (\ref{eq:schrodinger2}). Despite it is always symmetric, for convenience we have plotted the negative as well as the positive frequency part of the spectrum. In plot (a), one can clearly identify the two families of peaks composing the spectrum: the harmonic family (filled black triangles) and the fluorescence family (filled green squares), the filled red circles correspond  to the complex conjugate of this later. The harmonic family contains its own conjugate. The apparent complexity of the spectra at high field amplitudes, for instance the case shown in Fig. \ref{fig:two_spectra}(b), raises when the fluorescent family extends to negative frequencies, and its conjugate to positive. In this case each harmonic appears as surrounded by two satellite peaks, one belonging to the fluorescence family and the other to its complex conjugate.  Figure \ref{fig:full_spectra}(b) shows the same case as \ref{fig:two_spectra}(b), with the peaks labelled according to their correspondent family. As mentioned above, in general the intensity of the satellites around a particular harmonic is not the same, as the relative position of each satellite peak in the fluorescence family and its conjugate do not coincide.


\section{Stark Shift}

The physical interpretation of $\alpha_{0}$ in the fluorescence family can be found in the limit of low intensities $\chi \rightarrow 0$. As discussed before, in this limit the dipole evolution corresponds to the classical harmonic oscillator. This later problem is described only by two spectral contributions, one peak at the laser frequency $\omega$ and other at the natural frequency $\omega_0$ of the oscillator. As the field increases, the two-level system reveals its non-linearity, and natural and field frequency are mixed. Clearly, the harmonic spectrum rises from the mixture of the field frequency with itself, while the fluorescence family comes out of the mixture of the natural and field frequencies. Therefore, the central frequency $\alpha_0$ of the fluorescence family is to be interpreted as the effective transition frequency of the two-level atom, which is AC Stark shifted from $\omega_0$ as the field increases.

\par
Next we shall derive an approximated expression for the Stark shifted transition frequency. To do this, we evaluate the effective transition frequency $\alpha_0$ by solving $\det(M)=0$. For an arbitrary large truncation of $M$, a limiting exact solution for this quantity can be found numerically. If we are interested, however, in closed analytical estimations we should consider the regime of moderate coupling, for which the fluorescence family may be composed of several peaks but with a clear maximum at the central frequency $\alpha_0$. This situation allows us to neglect the influence of the peaks at the sides of $\alpha_0$ and to consider only the central term of matrix $M$ (i.e. an $1 \times 1$ truncation). The effective transition  given by $\det(M)=0$ reduces to $\Theta(\alpha_0)=0$, whose solutions are
\begin{eqnarray} \nonumber
\alpha_{0,\pm}^2={1 \over 2} \left[ \omega^2+\omega_0^2+2 |\chi|^2 \right. \hspace{3cm} \\ \left. \pm \sqrt{ (\omega^2+\omega_0^2+2 |\chi|^2)^2-4 \omega^2 \omega_0^2} \right] \; . \label{Stark_Shift}
\end{eqnarray}

\begin{figure}[htbp]
\centering\includegraphics[scale=0.48]{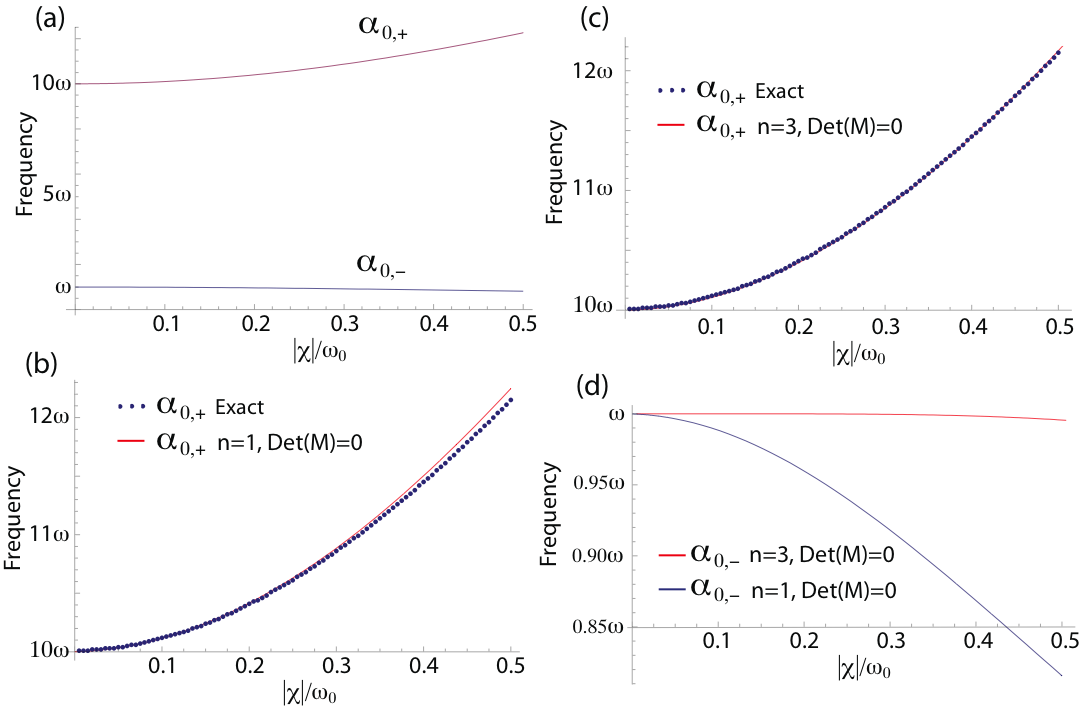}
\caption{\small (a) Plot of formula (\ref{Stark_Shift}) for $\omega/\omega_{0}=0.1$ versus the Rabi frequency, resulting from the truncation of $M$ to $1 \times 1$. $\alpha_{0,\pm}$ refer to the frequencies of the central peak in each of the two  families composing the dipole spectrum.  Since $\alpha_{0,-}$ corresponds to the harmonic family,  it is practically insensitive to the intensity field . (b) Test of the analytical solution $\alpha_{0,+}$ against  the exact numerical results from Eqs. (\ref{eq:schrodinger1}) and (\ref{eq:schrodinger2}), for $\omega/\omega_{0}=0.1$. The numerical data shows a good agreement with the analytical formula for the Stark shift although for large intensities small deviations appear. Parts (c) and (d) of this figure show the results for $\alpha_{0,\pm}$ resulting from the truncation of $M$ to $3 \times 3$. The results are obtained numerically from the equation $\det(M_{3 \times 3})=0$ and are compared with: (c) the exact numerical result also shown in part (b) of this figure, and (d) with the results for $\alpha_{0,-}$ from the $1 \times 1$ truncation shown in part (a) of this figure. }
\label{fig:Stark_Shift}
\end{figure}

The stark shifted transition frequency corresponds to $\alpha_{0,+}$ (which has the correct limit $\omega_0$ when $|\chi|$ tends to 0), while $\alpha_{0,-}$ converges to the laser frequency in the limit $|\chi|$ tending to 0. Therefore, the two solutions in  (\ref{Stark_Shift}) represent the two possible spectral families, harmonics and fluorescence, which are restricted to single peaks as a result of the $1 \times 1$ truncation of $M$. In the limit of small field intensities, and for $\omega_0 \ge 2 \omega$, Eq. (\ref{Stark_Shift}) is reduced to a Stark shift of $|\chi|^2 \omega_0/ (\omega_0^2-\omega^2)$, in correspondence with the form presented in \cite{marti05A}. 

Note that, even with the drastic truncation of $M$, the harmonic $\alpha_{0,-}$ is practically insensitive to the intensity field. In Fig. \ref{fig:Stark_Shift}(a) we plot Eq. (\ref{Stark_Shift}) for $\omega/\omega_{0}=0.1$ (taking only the positive frequencies), increasing the field from $\vert \chi \vert / \omega_{0} = 0.01$ to $\vert \chi \vert / \omega_{0} = 0.5$. The figure clearly shows that $\alpha_{0,-}$ remains practically constant as the intensity field increases, at variance with $\alpha_{0,+}$, which clearly depends on the field intensity. The comparison of the analytical solution $\alpha_{0,+}$ given by Eq. (\ref{Stark_Shift}) with the numerical results of the fluorescence peak computing Eqs. (\ref{eq:schrodinger1}) and (\ref{eq:schrodinger2}) give rise to a good match, see Fig. \ref{fig:Stark_Shift}(b). As expected, the approximated formula obtained by (\ref{Stark_Shift}) is practically exact at low field intensities but gradually deviates, at large intensities, from the exact numerically calculated solution.

As discussed before, a better accuracy can be found including the influence of the fluorescence peaks sorrounding the central one at $\alpha$. For instance, the nearest neighbors are included if we truncate $M$ to a $3 \times 3$ matrix. In this case, $\det(M)=0$ yields a polynomial of order 12 in $\alpha$, that cannot be solved analytically. However, the numerical solution of the problem can be attained easily in this case, and also for higher truncation orders. In Fig. \ref{fig:Stark_Shift}(c) we plot the Stark shift of the fluorescence transition and the points obtained by $\det(M_{3\times3})=0$. As expected, the accuracy of the results increases in this latter case compared with the order truncation $n=1$. We depict in Fig. \ref{fig:Stark_Shift}(d) $\alpha_0$ for the harmonic family in the $n=1$ and $n=3$ truncation order. Note that the dependence of this quantity with the field strength is smaller in the latter case, as the frequency of the harmonic should be independent of the intensity in the exact case (see the following section).


\section{The Harmonic Family}


In order to analyze the harmonic spectrum, we consider the matrix (\ref{eq:Matrix}) evaluated in  $\alpha_0 = \omega$. First, we demonstrate that in this case the condition ${\rm det} (M)=0$ is independent of the resonance frequency ($\omega_{0}$) or the laser intensity ($|\chi|$). Note from (\ref{eq:Phi}) that $\Phi_{+} (\alpha = \omega) = 0$ and $\Phi_{-}(\alpha=-\omega)=0$. These two zeros in $M$ provide a supplementary symmetry that allows us to reduce the complexity of the matrix to the following structure
\begin{widetext}
\begin{equation} M(\alpha)=
\left( \begin{array}{c|cc|cc}
   A\, (-\omega)_{\frac{n-1}{2}\times\frac{n-1}{2}} &   \Phi_{+}(-3\omega) &  &  & \\ \hline
   0 &  \Theta(-\omega) &  \Phi_{+}(-\omega) & & \\
    & \Phi_{-}(\omega)&  \Theta(\omega)& 0 &  \\ \hline
    &  & \Phi_{-}(3\omega)  & A\, (\omega)_{\frac{n-1}{2}\times\frac{n-1}{2}} &  \\
\end{array} \right) \; , \label{eq:Matrix_2}
\end{equation}
\end{widetext}
where $A\, (\pm \omega)$ are  ${\frac{n-1}{2}\times\frac{n-1}{2}}$ submatrices ($n$ is the truncation of the matrix $M$), which depends on the laser frequency $\omega$, but also on the resonance frequency and the laser intensity.  Now, starting from the new form (\ref{eq:Matrix_2}) of the truncated matrix $M$, we can calculate its determinant as
\begin{eqnarray} \nonumber
{\rm det} [M] = {\rm det} [A(\omega)] \cdot {\rm det} [A(-\omega)]  \times \hspace{1cm} \\ \left[\Theta(\omega)\Theta(-\omega)- \Phi_{+}(-\omega)  \Phi_{-}(\omega)\right]=0 \; ,
\end{eqnarray}
since $\left[\Theta(\omega)\Theta(-\omega)- \Phi_{+}(-\omega)  \Phi_{-}(\omega)\right]=0$, see equations (\ref{eq:Theta}) and (\ref{eq:Phi}). Therefore the location of the harmonic familiy is independent of the laser intensity.
Let us now study the different aspects of the harmonic spectrum.

\subsection{Harmonic ratios and relative phases}

\begin{figure}[htbp]
\centering\includegraphics[scale=0.4]{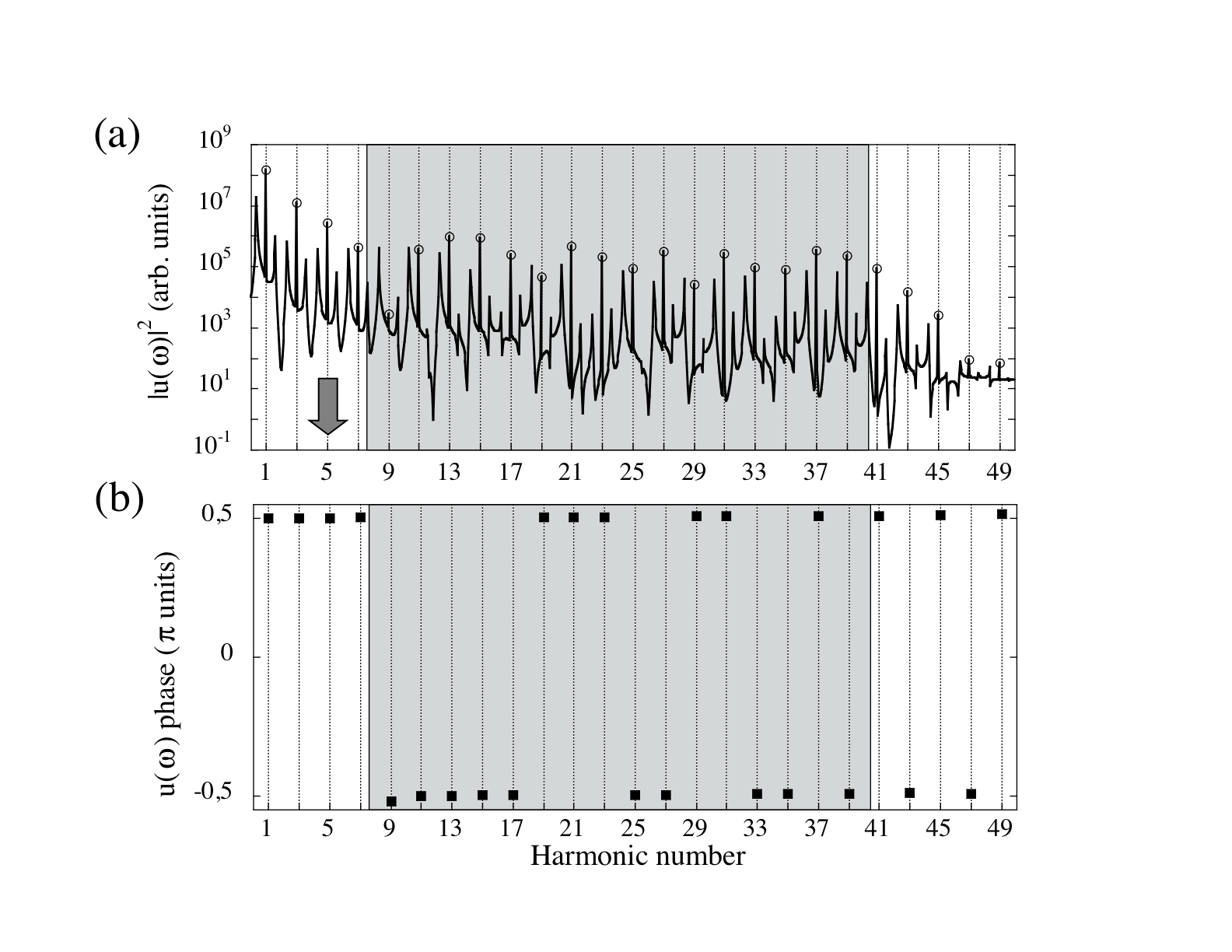}
\caption{\small Dipole spectral intensities (a) and phases (b) for the strongly driven case  with $|\chi|/\omega_0=4$ and $\omega/\omega_0=0.2$. The harmonic peak family is highlighted with open circles. The shadowed box encloses the harmonic plateau structure. The limits of this box have been defined using the expressions (\ref{eq:onset}) and (\ref{eq:cutoff}) for the plateau on-set and cut-off frequencies. The arrow points to the plateau's on-set frequency in the weak driving limit, equal to $\omega_0$.}
\label{fig:phases}
\end{figure}

\begin{figure}[htbp]
\centering\includegraphics[scale=0.4]{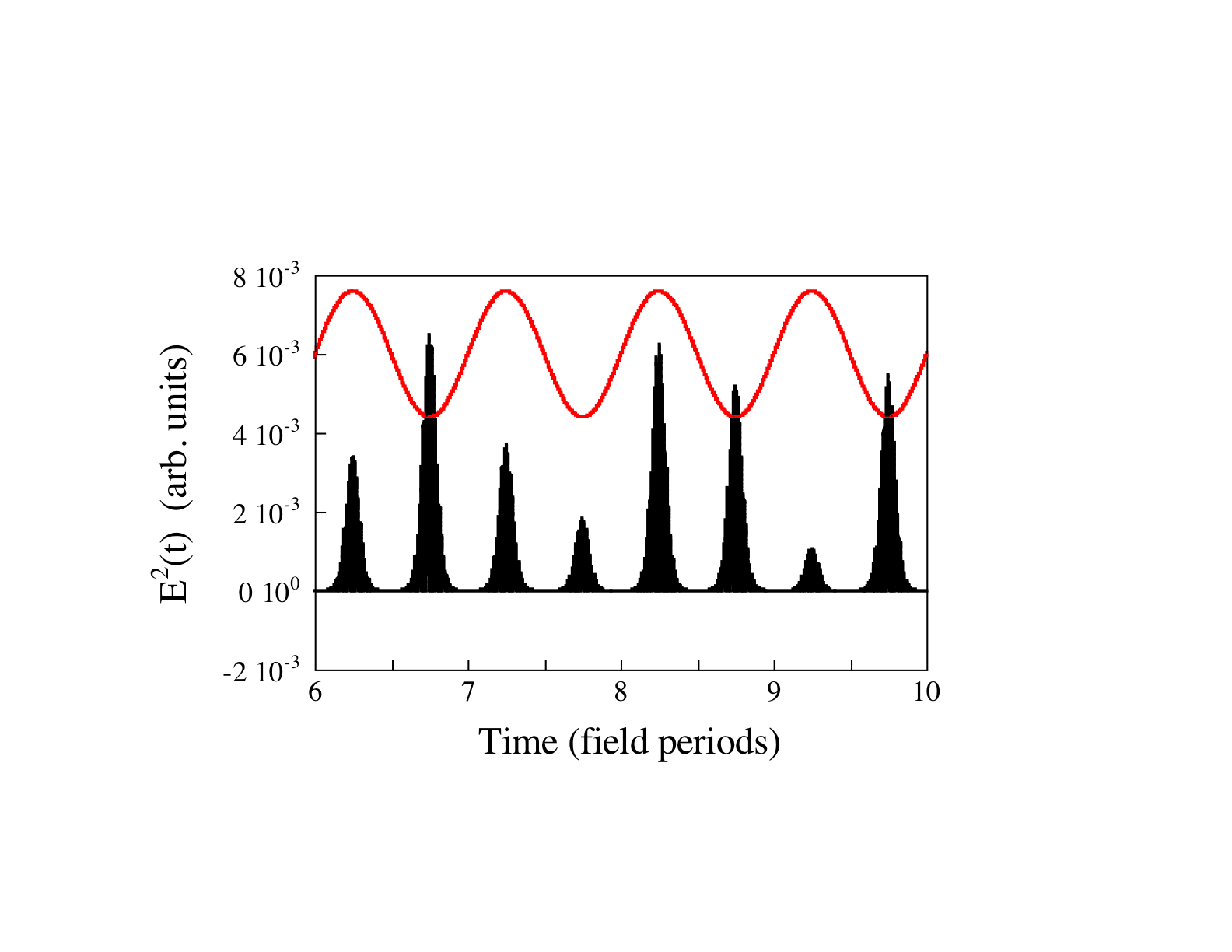}
\caption{\small Black filled curve: detail of the inverse Fourier transform of the higher frequency part of the dipole spectrum shown in figure \ref{fig:phases}. Frequencies below $34 \omega$ are filtered out. Only the field envelope is represented. Red line: Sketch of the amplitude of the driving field at the same time interval.}
\label{fig:attos}
\end{figure}

Coming back to Eq. (\ref{eq:system}), and defining the ratio between neighboring harmonics as $Z(\alpha)=u(\alpha)/u(\alpha-2\omega)$, we have the following relation 
\begin{equation}
Z(\alpha)=- {\Phi_{-}(\alpha) \over \Theta(\alpha)+ \Phi_+(\alpha) Z(\alpha+2\omega)} \; .
\label{eq:ratios}
\end{equation}
This recursive expansion is exact and, therefore, reproduces accurately the relative weights between the peaks of the harmonic and fluorescence spectral families \cite{plaja92A}. While the ratios between the harmonics inside the plateau region form a complex sequence, it is not so in the regions outside the plateau, where the harmonics decrease monotonically with the frequency, and the relative ratio $Z(\alpha)$ is a small quantity. In such regions we may approximate
\begin{equation}
Z(\alpha)\simeq- {\Phi_{-}(\alpha) \over \Theta(\alpha)} \; .
\label{eq:aprox_ratios}
\end{equation}
The spectral region with frequencies below the plateau is characterized by the conditions $\alpha>2 \omega$ and $\alpha< \omega_0$, yielding $\Theta(\alpha)>0$ and $\Phi_{-} \propto  \exp(-2i\phi)$ . Consequently $Z(\alpha) \propto - \exp(-2i\phi)$ and the relative phase between consecutive harmonics before the plateau is $ \exp \left[ -2i\phi+ \pi \right]$. On the other hand, the spectral region above the plateau is characterised by  $\alpha$ arbitrarily large, therefore, $\Theta(\alpha)<0$ and $\Phi_{-} \propto \exp(-2i\phi)$. Consequently, the relative phase between consecutive harmonics after the plateau is $ \exp(-2i\phi)$. Inside the plateau region, the phase distribution is in general more complex and can only be determined using the recurrent relation (\ref{eq:ratios}) analogously as it was done in \cite{plaja92A} for the harmonic intensities. Figure \ref{fig:phases} shows the harmonic spectrum and phases computed numerically from Eqs. (\ref{eq:blochu}) to (\ref{eq:blochw}) in the strong coupling case ($|\chi|/\omega_0=4$ and $\omega/\omega_0=0.2$). The plateau region is shaded in grey (for a detailed discussion of the plateau's limits see below). As stated before, the field phase has been taken as $\phi=-\pi/2$ and, therefore, the above discussion predicts a relative phase between consecutive harmonics of $2\pi$ before the plateau and $\pi$ after the plateau, in accordance with the numerical results shown in figure \ref{fig:phases}. Note that the above phase relations hold also approximately in the extreme parts of the plateau. This is a relevant aspect, as it implies that the harmonics near the plateau's cut-off are approximatelly phase locked. This spontaneous locking is also found in harmonic generation with ionizing systems, and implies the possibility of synthesis of attosecond pulses from the Fourier synthesis of the harmonics near the cut-off.  This possibility is analyzed in figure \ref{fig:attos} in which we present the time evolution of the dipole resulting from the inverse Fourier transform of the spectral components with frequencies above $34 \omega$, for the case shown in figure \ref{fig:phases}. The black filled curves show the squared envelope of the dipole, resulting in a train of pulses with durations well below the optical period. The irregularity (no periodicity) of the squared envelope of the dipole in Fig. \ref{fig:attos}  is because of the resonant family contribution. 
 
 \subsection{Plateau's on-set and cut-off frequencies}
 
We can have an approximated idea for the plateau's extension by finding its limiting frequencies. To do this we will simply consider as a reasonable estimation the fact that the neighboring harmonics in the plateau have similar intensities: $|u(\alpha-2\omega)| \simeq  |u(\alpha)| \simeq  |u(\alpha+2\omega)| $. The plateau on-set can be obtained using this condition together with  the relative phase for the lower frequency harmonics derived above: $\exp(-2i\phi+\pi)$. For this case Eq. (\ref{eq:system}) yields to
\begin{equation}
\Phi_-(\alpha) e^{+2i\phi}- \Theta(\alpha) + \Phi_+(\alpha) e^{-2i\phi}=0 \; ,
\end{equation}
which has a solution
\begin{equation}
\alpha^2= {1 \over 2} \left[  \omega_0^2 + \omega^2 + \sqrt{ \left( \omega_0^2 - \omega^2 \right)^2+ 16 \omega^2 |\chi|^2} \right] \; .
\label{eq:onset}
\end{equation}
In the limit  $\omega \rightarrow 0$, the plateau's on-set is approximately $\omega_0$ which is the estimation in  \cite{kapla94A}.

The value of the frequency at the plateau's cut-off can be inferred imposing the phase condition for the higher frequency harmonics: $\exp(-2i\phi)$. We have then
\begin{equation}
\Phi_-(\alpha) e^{+2i\phi}+ \Theta(\alpha) +\Phi_+(\alpha) e^{-2i\phi}=0 \;,
\end{equation}
which has a solution
\begin{equation}
\alpha^2=\omega_0^2+4  |\chi|^2 \; ,
\label{eq:cutoff}
\end{equation}
in coincidence with \cite{kapla94A} and also converges to  \cite{ivano93B} in the limit of $\omega_0 \rightarrow 0$. For large field intensities, the plateau extends linearly with the field amplitude in contrast with the case of    high-order harmonic generation in ionizing systems, where the cut-off is proportional to the intensity.

\begin{figure}[htbp]
\centering\includegraphics[scale=0.35]{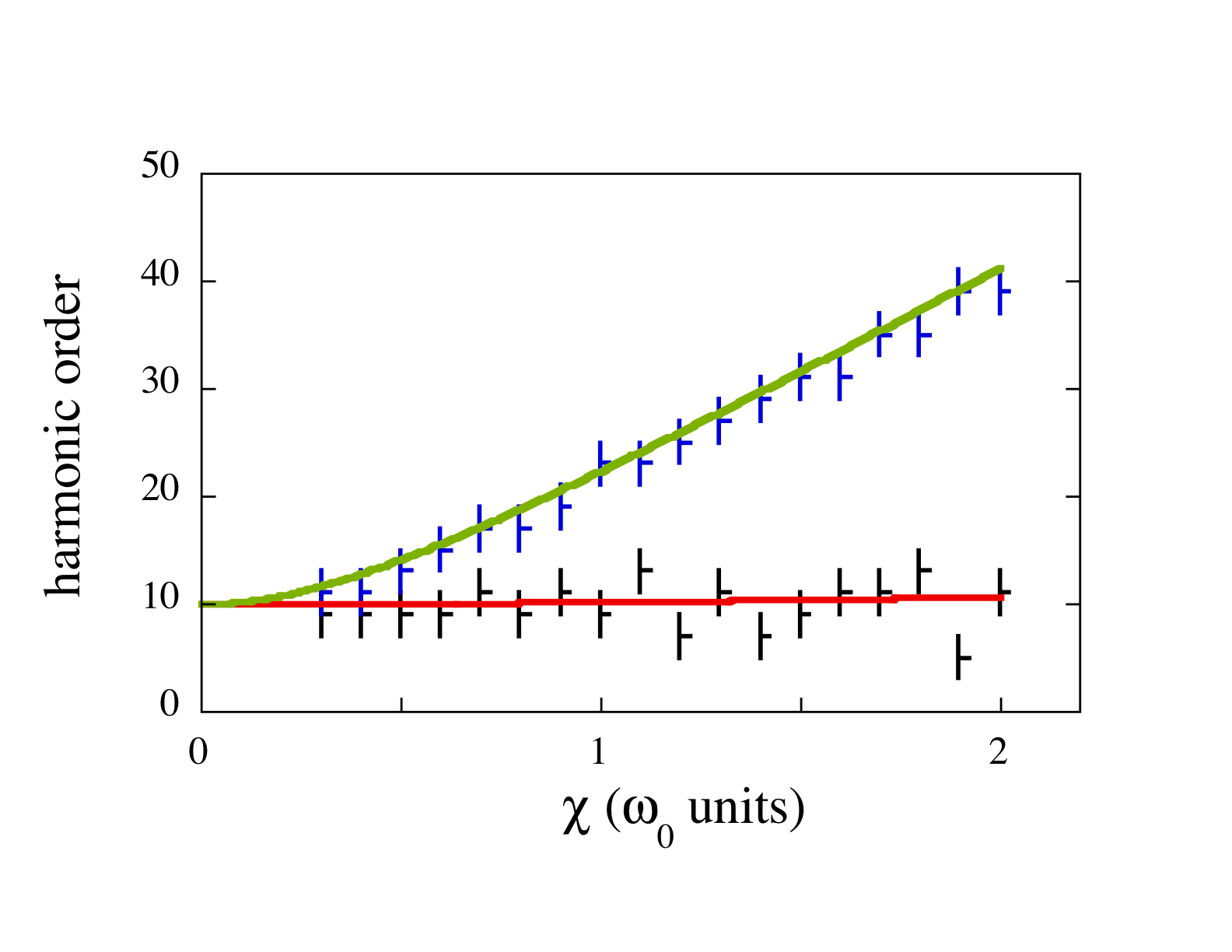}
\vspace*{-1.0 cm}
\caption{\small Frequencies for the plateau's on-set (red line) and cut-off (green line) derived from Eqs. (\ref{eq:onset}) and (\ref{eq:cutoff}) for the case $\omega/\omega_0=0.1$. The visual estimation from the numerically computed spectra are plotted with black and blue points, we have added an error bar of $\pm 2 \omega$ as an error estimation of the method.}
\label{fig:on_off}
\end{figure}

Figure \ref{fig:on_off} shows the results of Eqs. (\ref{eq:onset}) and (\ref{eq:cutoff}) for the case $\omega/\omega_0=0.1$ and different field amplitudes. Superimposed to this, we plot the values of the harmonic number for the plateau's on-set and cut-off extracted from the numerical integration of Eqs. (\ref{eq:schrodinger1}) and (\ref{eq:schrodinger2}). For the latter case the plateau's on-set remains practically constant and equal to $\omega_0$, however for larger photon frequencies, the departure between Eq.  (\ref{eq:onset}) and the low coupling limit $\omega_0$ becomes more apparent. For instance in the case plotted in figure \ref{fig:phases} the plateau's onset frequency evaluated from Eq.  (\ref{eq:onset}) (corresponding to the lower limit of the shadowed box) is about $2 \omega$ above $\omega_0$ (pointed out by an arrow).

\section{Physical intepretation: the adiabatic regime.}

\begin{figure}[htbp]
\centering\includegraphics[width=7cm]{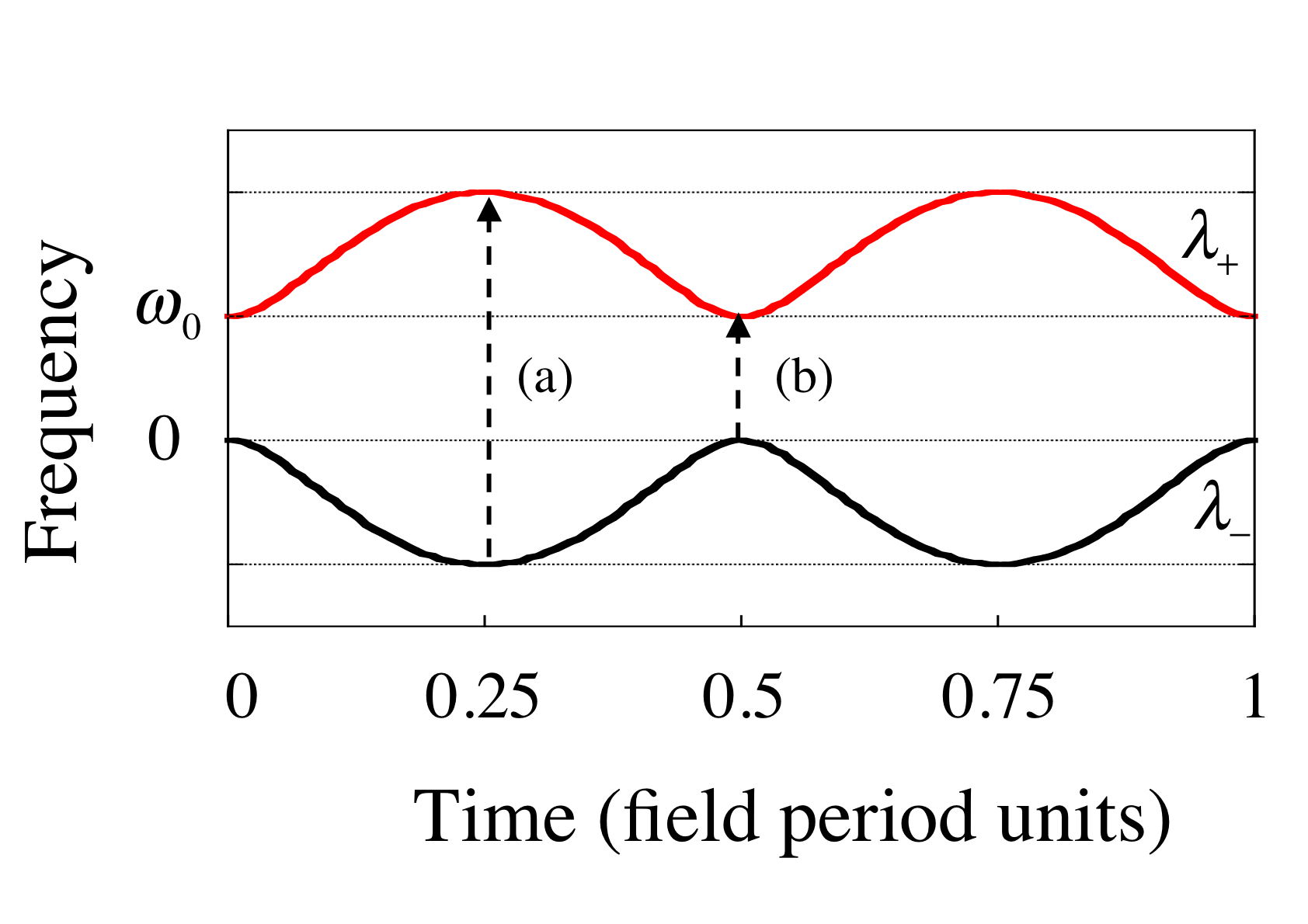}
\caption{\small Scheme of the instantaneous eigenenergies of the two level atom during one driving field period. In the adiabatic limit, the maximum (a) and minimum (b) level differences define the energies of the plateau on-set and cut-off respectively. }
\label{fig:dressed}
\end{figure}

The above formulas have a clear physical interpretation if we consider the limit of small driving frequencies, $\omega\ll\omega_0$ and $\omega\ll|\chi|$  \cite{conej96A,figue02A} . In this case we can consider the instantaneous eigenstates of the time dependent Hamiltonian as physically meaningful. The diagonalization of (\ref{eq:schrodinger1}) and (\ref{eq:schrodinger2}) gives the eigenenergies
\begin{equation}
\lambda_{\pm}={1 \over 2} \left( \omega_0 \pm \sqrt{\omega_0^2+ 4 F^2(t)} \right) \; ,
\label{eq:transition}
\end{equation}
with $F(t)= Re \left \{ \chi \exp(-i \omega t)\right \}$, oscillating harmonically. Figure \ref{fig:dressed} shows an schematic plot of these two eigenenergies along a laser cycle. The instantaneous transition energy  is given by $\lambda_+-\lambda_-= \sqrt{\omega_0^2+ 4 F^2(t)}$ which oscillates between $\omega_0$ (at times when $F(t)=0$) and $\sqrt{\omega_0^2+ 4 |\chi|^2}$  (at the maximum amplitude of $F(t)$). These two values coincide with the harmonic plateau on-set and cut-off at the limit $\omega \rightarrow 0$ derived in the above section, see Eqs. (\ref{eq:onset}) and (\ref{eq:cutoff}). This suggests that the plateau structure of the two-level dipole spectrum is originated by the transitions between the field dressed states. This mechanism is different from the strong-field scenario, where the same structure is originated by the high energy radiation emitted from the recollision between an ionized electron with the parent atom. A fundamental difference between these two cases also appears in the identification of the moment where the higher frequency harmonics are emitted. In the two-level case, Eq. (\ref{eq:transition}), these harmonics correspond to the maximum of $F^2$ and, therefore, to the maxium of the driving field, while in real atoms the ionised electron recollides with the parent atom at times where the driving field is near zero.  

\section{Conclusions}
We present a formalism to fully describe the dipole spectrum of a two-level system excited by a non-resonant intense electromagnetic field. The derived formalism beyond the rotating wave approximation is general within the electric dipole approximation for monochromatic fields. This new approach allows to perform a fundamental decomposition of the complex dipole spectrum structure into two families: the harmonic frequencies of the driving field and the field-induced nonlinear fluorescence frequencies. Moreover, it provides analytical expressions for the non-resonant Stark shift for the fluorescence family and of the on-set and cut-off limits for the harmonic family. Within this formalism, we predict the generation of pulses with lower duration than the period of the driving field by selectively filtering out the frequencies of the emitted radiation, in close analogy with the attosecond pulse generation in the case of ionizing electromagnetic fields \cite{brabe00A}. 
\par
It is important to note that the presented formalism is an optimal tool to treat the burgeoning new physical scenarios where the two-level system approximation is applied, such as harmonic generation in molecules or dipolar two-level transitions in nuclei. Therefore, our approach is suitable to theoretically address near future experiments in the field.\\

\par


\section{Acknowledgments}

We acknowledge G. Orriols for fruitful discussions. We also acknowledge support by the Spanish Ministerio de Ciencia e Innovaci\'on, FIS2007-29091-E, FIS2008-02425, FIS2006-04151, FIS2009-09522, Consolider programs SAUUL and QOIT under contracts CSD2007-00013 and CSD2006-00019, respectively, and by the Catalan Government under contract SGR2009-00347 and by the Junta de Castilla y Le\'on (SA146A08).

\end{document}